\documentstyle[epsf,twocolumn,seceq]{jpsj}

\def\runtitle{Magnetic Properties of a Pressure-induced Superconductor UGe$_2$}
\def\runauthor{Naoyuki {\sc TATEIWA} {\it et al}}

\title
{
Magnetic Properties of a Pressure-induced Superconductor UGe$_2$ 
}

\author
{ 
Naoyuki {\sc TATEIWA}\footnote{E-mail:tateiwa@djebel.mp.es.osaka-u.ac.jp}, Katsumi HANAZONO, Tatsuo C KOBAYASHI$^{1}$, Kiichi AMAYA, Tetsutaro INOUE$^{1}$, Koichi KINDO$^{1}$, Yoshihiro KOIKE$^{2,4}$, Naoto METOKI$^{2}$, Yoshinori HAGA$^{2}$, Rikio SETTAI$^{3}$ and Yoshichika \=ONUKI$^{3}$
}

\inst
{
Department of Physical Science, Grasuate School of Engineering Science, Osaka University, Toyonaka, Osaka 560-8531, Japan\\
$^1$Research Center for Materials Science at Extreme Conditions, Osaka University, Toyonaka, Osaka 560-8531, Japan\\
$^2$ Advanced Science Research Center, Japan Atomic Energy Research Institute, Tokai-mura, Naka-Gun, Ibaraki, 319-1195, Japan\\
$^3$ Department of Physics, Graduate School of Science, Osaka University, Toyonaka, Osaka 560-0043, Japan\\
$^4$ Japan Science and Technology Corporation, Kawaguchi, Saitama 332-0012, Japan\\
}

\recdate
{
\today
}

\abst
{
We performed the DC-magnetization and neutron scattering experiments under pressure {\it P} for a pressure-induced superconductor UGe$_2$. We found that the magnetic moment is enhanced at a characteristic temperature  {\it T}$^{\,*}$ in the ferromagnetic state, where {\it T}$^{\,*}$  is smaller than a Curie temperature {\it T}$_{\rm C}$. This enhancement becomes remarkable in the vicinity of {\it P}$_{\rm C}^{\,*}$ = 1.20 GPa, where {\it T}$^{\,*}$  becomes 0 K and the superconducting transition temperature {\it T}$_{\rm SC}$ shows a maximum. The characteristic temperature {\it T}$^{\,*}$ , which decreases with increasing pressure, also depends on the magnetic field.
}

\kword
{
UGe$_2$, Ferromagnetism, Superconductivity
}

\begin{document}
\sloppy
\maketitle

  UGe$_2$ is known as a metallic ferromagnet with a Curie temperature of {\it T}$_{\rm C}$ = 52 K at ambient pressure.~\cite{rf:1} The magnetization is extremely anisotropic, with the easy {\it a}-axis in the orthorhombic crystal structure. The ordered moment is 1.4 $\mu_{\rm B}$/U well below {\it T}$_{\rm C}$.~\cite{rf:2} Application of pressure suppresses {\it T}$_{\rm C}$, and finally {\it T}$_{\rm C}$ becomes zero at the critical pressure {\it P}$_{\rm C}$ of 1.6-1.7 GPa.~\cite{rf:3} 

  Recently superconductivity was observed under pressure in UGe$_2$.~\cite{rf:4,rf:5,rf:6}  Superconductivity appears in the pressure range from 1.0 to 1.6 GPa. In other words the present superconductivity disappears at $P > P_{\rm C}$, namely in the paramagnetic state. The superconducting transition temperature {\it T}$_{\rm SC}$ shows a maximum of 0.8 K around 1.2 GPa where the ferromagnetic state is still stable with {\it T}$_{\rm C}$=32K.~\cite{rf:4,rf:5,rf:6}  The heat capacity anomaly due to the superconducting transition was found at 1.13 GPa, which indicates the bulk property of superconductivity.~\cite{rf:6}  From the elastic neutron scattering experiment under pressure, the ferromagnetic moment of about 1.0 $\mu_{\rm B}$/U was found even in the superconducting state.~\cite{rf:5}  It is thus concluded that superconductivity and ferromagnetism coexist below {\it T}$_{\rm SC}$.

   In UGe$_2$ another characteristic anomaly exists at {\it T}$^{\,*}$ in the ferromagnetic state below {\it T}$_{\rm C}$.~\cite{rf:3,rf:5,rf:6} This anomaly, which was observed on the resistivity, becomes sharper with applying pressure.Å@Application of the pressure also suppresses {\it T}$^{\,*}$, and finally {\it T}$^{\,*}$ becomes 0 K at the another critical pressure {\it P}$_{\rm C}^{\,*}$ = 1.2 GPa, which is smaller than $P_{\rm C}$. The superconducting transition temperature {\it T}$_{\rm SC}$ shows a maximum around {\it P}$_{\rm C}^{\,*}$.~\cite{rf:5,rf:6} Å@In the pressure just above {\it P}$_{\rm C}^{\,*}$, the characteristic transition {\it T}$^{\,*}$ is induced by the magnetic field and then the upper critical field {\it H}$_{\rm C2}$ is enhanced.~\cite{rf:5} Similar enhancement of {\it H}$_{\rm c2}$ has been confirmed in our resistance measurement.~\cite{rf:7} Therefore it is natural to consider that the critical fluctuation related to the transition at {\it T}$^{\,*}$ may be a driving force for the superconducting pairing in UGe$_2$. The previous elastic neutron scattering experiment under high pressure also revealed that the magnetic scattering intensity was slightly enhanced below {\it T}$^{\,*}$ at 0.89 GPa.~\cite{rf:5}  Cambridge and Grenoble groups suggested the existence of the CDW/SDW state below {\it T}$^{\,*}$ because the nesting was expected from the topology of the Fermi surface.~\cite{rf:4,rf:5,rf:8,rf:9}
  
 In this study, the magnetic properties under high pressure were investigated by the magnetization and elastic neutron scattering experiments in order to investigate the nature of the characteristic transition which occurs at {\it T}$^{\,*}$.

 A single crystal of UGe$_2$ was grown by the Czochralski pulling method in a tetra-arc furnace as described in Ref. 6. The purity of the starting materials was 99.98 \% for U and 99.999 \% for Ge. The ingot was annealed at 800 $^{\circ}$C in a high vacuum of 5Å~10$^{-11}$ torr for 7 days. As for the present sample, the residual resistivity $\rho_{0}$ and the residual resistivity ratio RRR (= $\rho_{\rm\, RT}/\rho_{0}$ , $\rho_{\rm \, RT}$: the resistivity at room temperature) were 0.26 $\mu\Omega$cm and 600, respectively, at ambient pressure, indicating a high-quality sample.
  \begin{figure}
 \begin{center}
  \epsfxsize=8.5cm
  \epsfbox{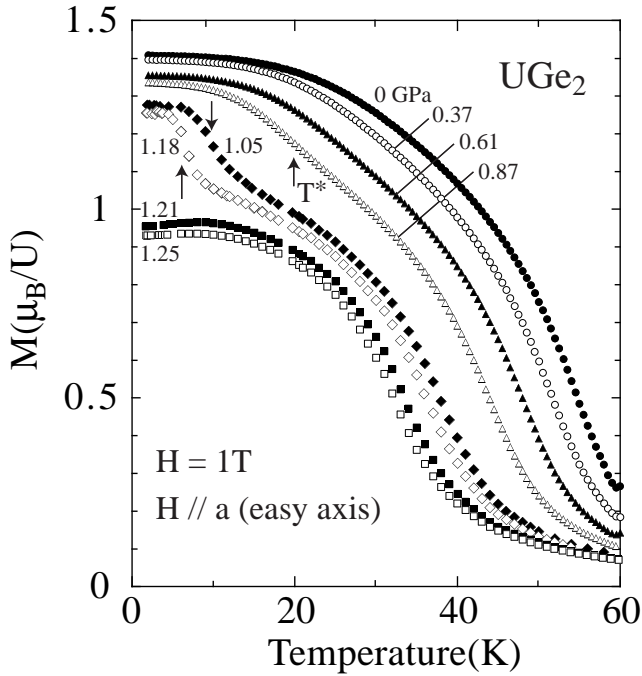}
 \end{center}
\caption{Temperature dependence of the magnetization at several pressures for the field of 1 T along the {\it a}-axis in UGe$_2$  }
\label{fig:1}
\end{figure}

\begin{figure}
 \begin{center}
  \epsfxsize=8.5cm
  \epsfbox{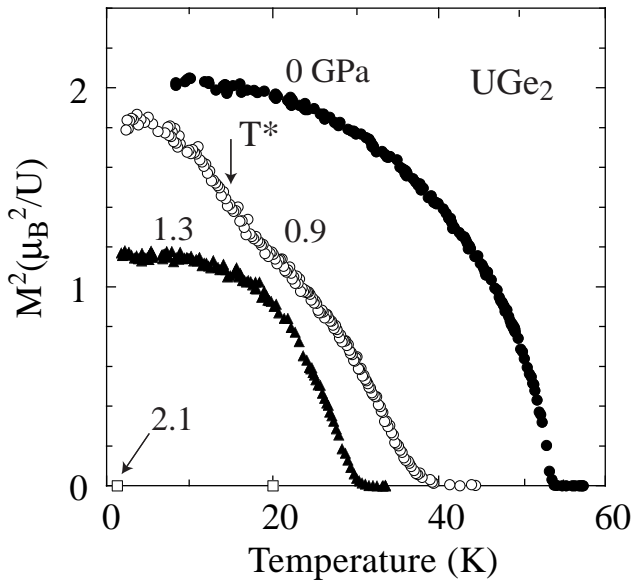}
 \end{center}
\caption{Temperature dependence of the squared magnetic moment measured by the (001) Bragg intensity in the neutron scattering experiment for UGe$_2$   }
\label{fig:2}
\end{figure}
 The magnetization measurement was carried out by using a SQUID magnetometer. Pressure was applied by utilizing a Cu-Be piston-cylinder cell with a Daphne oil (7373) as a pressure-transmitting medium. The pressure value was determined by the superconducting transition temperature  {\it T}$_{\rm SC}$ of lead. The lead chip was kept apart from the UGe$_2$ sample to avoid the field effect of  {\it T}$_{\rm SC}$ from the ferromagnetism of the sample. The field effect was negligibly small, which was checked at ambient pressure. 
\begin{figure}
 \begin{center}
  \epsfxsize=8.5cm
  \epsfbox{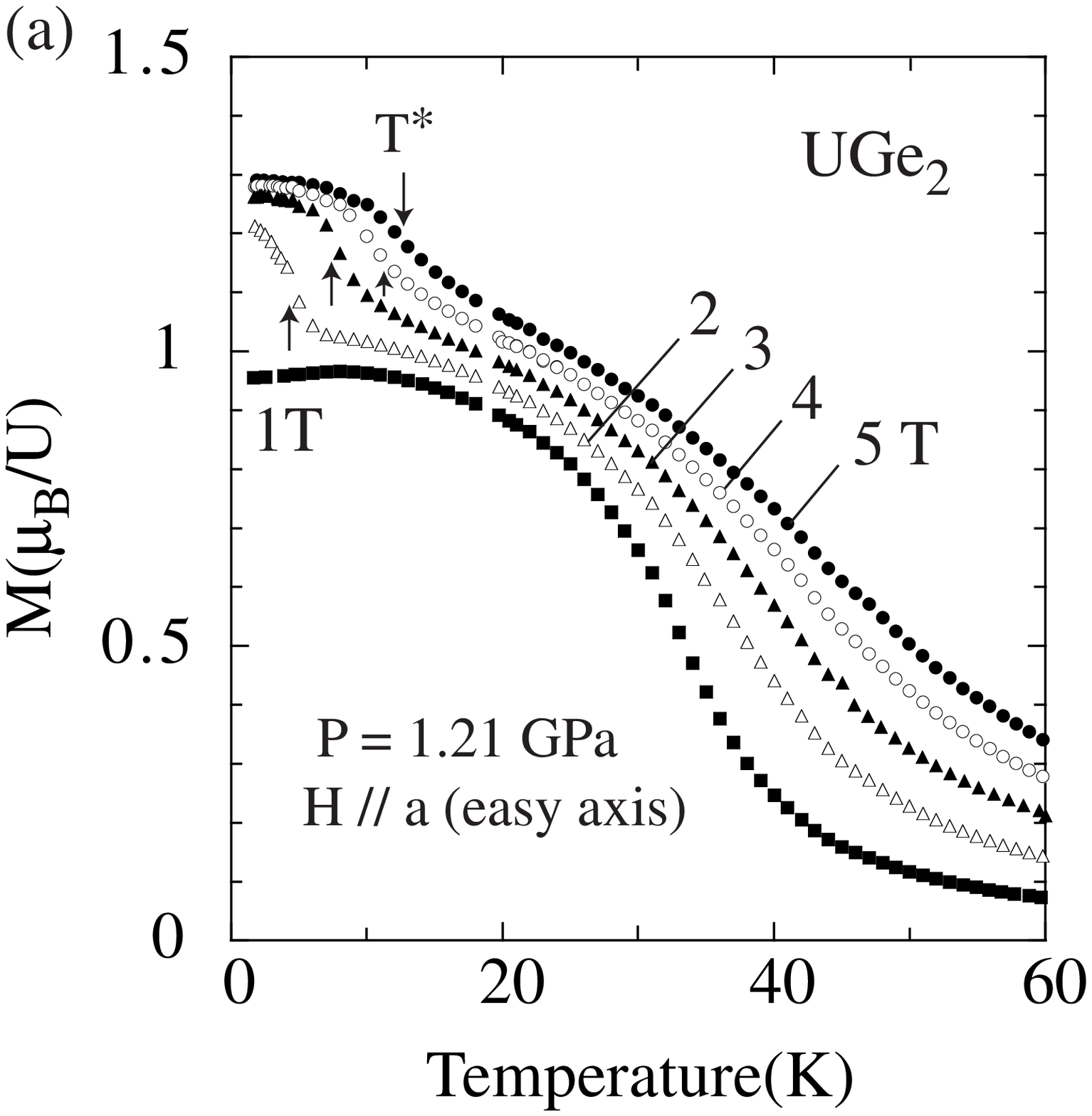}
 \epsfxsize=8.5cm
  \epsfbox{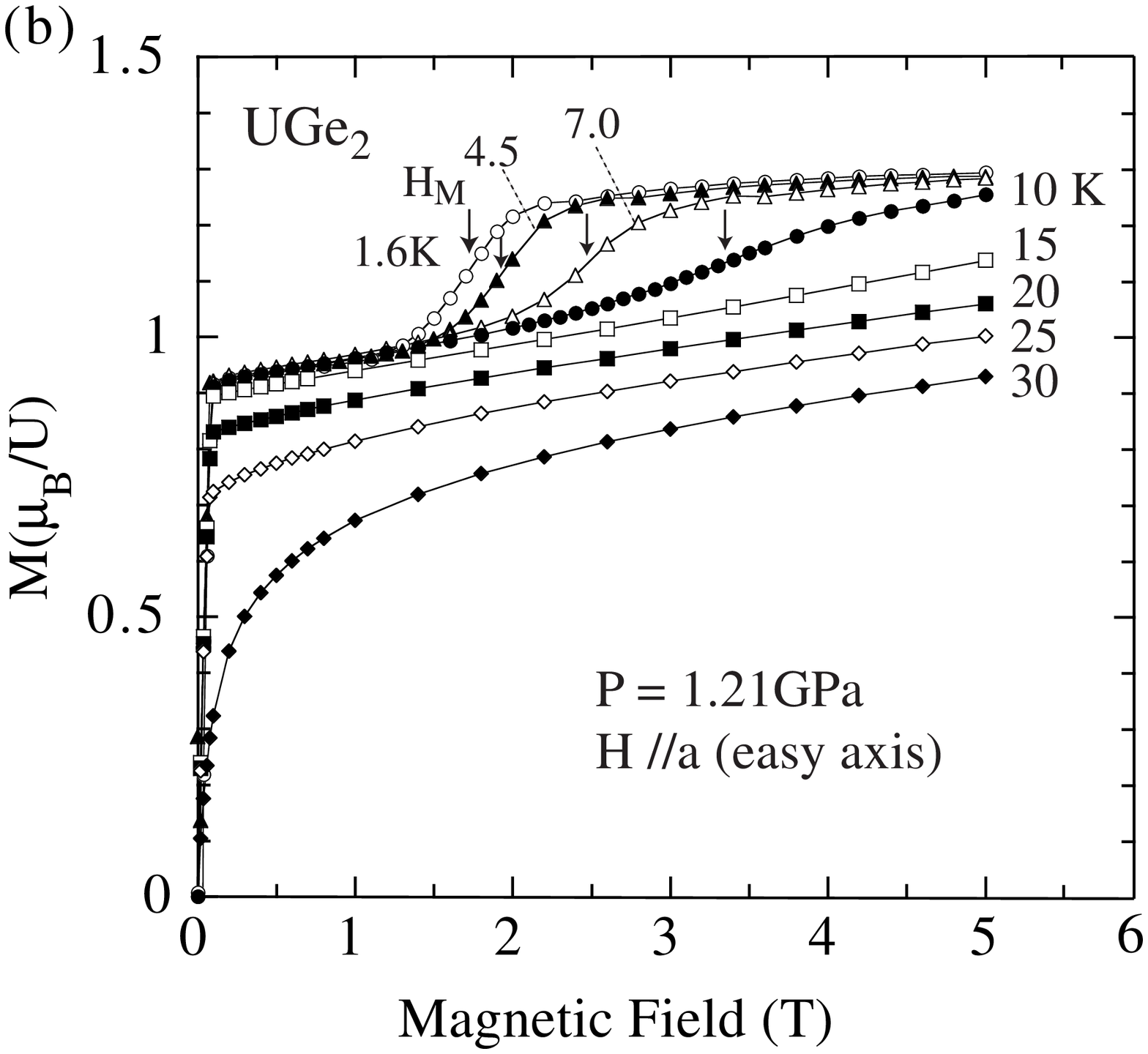}
 \end{center}
\caption{
(a) Temperature dependence of the magnetization at 1.21 GPa under various magnetic fields along the {\it a}-axis and (b) the magnetization curves at several temperatures UGe$_2$  }
\label{fig:4}
\end{figure}

 The neutron scattering experiment was carried out using a thermal and cold triple-axis-spectrometers installed at the research reactor JRR-3M in Japan Atomic Energy Research Institute (JAERI). Hydrostatic pressure was applied with use of a Macwhan-type cell. Fluorinert was sealed with a sample as a pressure transmitting medium. The pressure was applied at room temperature and calibrated at 4.2 K by the lattice constant of NaCl encapsulated with a sample. The sample was mounted with the [100] ({\it a}-axis) and [001] ({\it c}-axis) directions parallel to the scattering plane.
\begin{figure}
 \begin{center}
  \epsfxsize=8.5cm
  \epsfbox{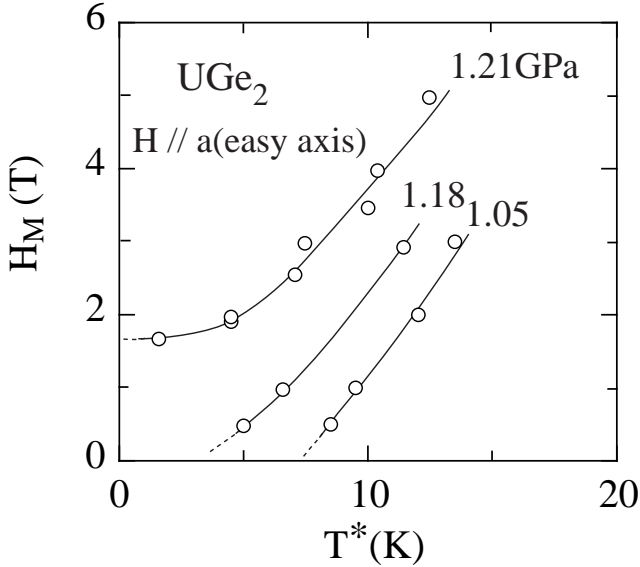}
 \end{center}
\caption{{\it T}$^{\,*}$ vs. {\it H}$_{\rm M}$ relation in the vicinity of {\it P}$_{\rm C}^{\,*}$ for {UGe$_2$} }
\label{fig:4}
\end{figure}

 Figure 1 shows the temperature dependence of the magnetization, ${\it M}({\it T})$, in the magnetic field of 1 T along the {\it a}-axis (easy axis). The result at ambient pressure agrees well with the previous report.~\cite{rf:1} In the pressure range close to {\it P}$_{\rm C}^{\,*}$, namely at {\it P} = 1.05 and 1.18 GPa, an abrupt step-like increase of the magnetization is found around {\it T}$^{\,*}$.~\cite{rf:8} 
The characteristic temperature {\it T}$^{\,*}$, defined by the peak of -$dM/dT$, is shown by arrows in Fig. 1. The magnetization is enhanced by about 20 \% at 1.05 and 1.18 GPa. The characteristic feature, which occurs at {\it T}$^{\,*}$, is found clearly in the vicinity of {\it P}$_{\rm C}^{\,*}$.In other words, at pressures far below {\it P}$_{\rm C}^{\,*}$, the anomaly becomes broader, as seen at {\it P} = 0 - 0.61 GPa. This feature is consistent with the resistance measurement.~\cite{rf:6} Namely, the corresponding anomaly, which is observed as a kink of resistivity in the vicinity of {\it P}$_{\rm C}^{\,*}$, becomes a shoulder at low pressures. Above {\it P}$_{\rm C}^{\,*}$ such an increase of the magnetization is not observed, and the magnetic moment at the lowest temperature decreases drastically to 0.92 $\mu_{\rm B}$/U at $P$ = 1.25 GPa. These results also indicate that the critical pressure {\it P}$_{\rm C}^{\,*}$ is 1.20 GPa. 

   To observe a spontaneous ferromagnetic component directly, we performed the elastic neutron scattering experiment under several pressures at zero magnetic field.  Figure 2 shows the temperature dependence of the squared magnetic moment measured by (001) Bragg intensities. At ambient pressure the (001) Bragg intensity increases with decreasing temperature below {\it T}$_{\rm C}$ = 53 K. With applying pressure of about $P$ = 0.9 GPa, {\it T}$_{\rm C}$ was reduced to about 40 K.  A clear step-like increase in the magnetic Bragg intensity was observed below {\it T}$^{\,*}$ = 18 K. At $P$ = 1.3 GPa the anomaly with {\it T}$^{\,*}$ disappears completely. The Curie temperature {\it T}$_{\rm C}$ and a saturated magnetic moment at $P$ = 1.3 GPa are 30 K and 1.1 $\mu_{\rm B}$/U, respectively. This saturated magnetic moment is about 20 \% larger than that obtained from the magnetization at $P$ = 1.25 GPa. This difference may come from the ambiguity in the determination of the magnetic moment in the neutron scattering experiment due to the influence of extinction. At $P$ = 2.1 GPa we observed no ferromagnetic transition down to 1.3 K.
\begin{figure}
 \begin{center}
  \epsfxsize=8.5cm
  \epsfbox{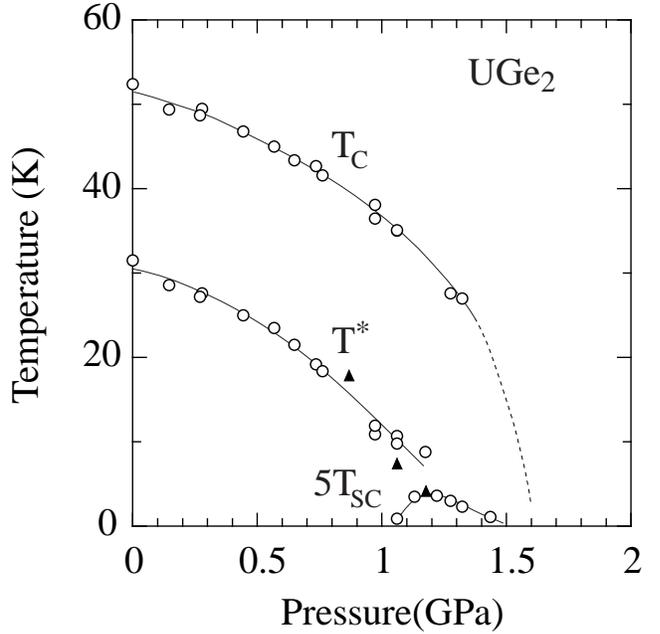}
 \end{center}
\caption{Phase diagram under pressure in UGe$_2$ obtained from the resistivity (denoted by circles) and the present magnetization (denoted by triangles) }
\label{fig:5}
\end{figure}

  Figure 3 (a) shows the temperature dependence of the magnetization at $P$ = 1.21  GPa under several magnetic fields. A remarkable step-like increase of the magnetization is observed above 2 T. It indicates that {\it T}$^{\,*}$ is induced by the magnetic field even above {\it P}$_{\rm C}^{\,*}$. The characteristic temperature {\it T}$^{\,*}$ shifts to a higher temperature with increasing the magnetic field. Similar increase of {\it T}$^{\,*}$ with increasing field is observed in the pressure range below {\it P}$_{\rm C}^{\,*}$. It is noted that a broad maximum of the magnetization exists around 10 K in the magnetic field of 1 T at $P$ = 1.21GPa. However the magnetization at $P$ = 1.25 GPa did not show such a feature as shown in Fig. 2. Therefore it is suggested that this broad maximum is characteristic in the magnetization just above {\it P}$_{\rm C}^{\,*}$. 

    Figure 3 (b) shows the magnetization curve at $P$ = 1.21 GPa. At the lowest temperature of 1.6 K, a step-like increase of the magnetization is observed, indicating a metamagnetic transition at {\it H}$_{\rm M}$ =1.7 T, where {\it H}$_{\rm M}$ is defined by the peak of $dM/dH$. Such an anomaly was observed in the AC susceptibility and magnetoresistance measurements.~\cite{rf:5} The present DC magnetization measurement reveals that the magnetic moment is enhanced by about 30 \% in this metamagnetic transition. The metamagnetic transition indicates no hysteresis, suggesting that this metamagnetic transition may not be a first order phase transition but may possibly be a second order one. 

  From these results, a {\it T}$^{\,*}$ vs. {\it H}$_{\rm M}$ relation was obtained at several pressures, as shown in Fig. 4. In the pressure region below {\it P}$_{\rm C}^{\,*}$ = 1.20 GPa, {\it H}$_{\rm M}$ increases almost linearly as a function of temperature. On the other hand, at $P$ = 1.21 GPa, there exists a threshold of the magnetic field {\it H}$_{\rm M} \sim 1.7 {\it T}$ in this transition. 

  With these transition temperatures, a temperature-pressure phase diagram of UGe$_2$ is obtained, as shown in Fig.5. The transition temperature denoted by circles were determined by the resistivityÅ@measurement, where {\it T}$^{\,*}$ was defined by the peak of $d{\rho} /dT$.~\cite{rf:6} The characteristic {\it T}$^{\,*}$ obtained in the present DC-magnetization measurement, defined by the interpolation of {\it T}$^{\,*}$ in the magnetic field to zero magnetic field, were denoted by filled triangles. It is noted that the step-like increase of the magnetization occurs below the temperature where the resistivity shows anomaly especially near {\it P}$_{\rm C}^{\,*}$.
 
  It is found in this study that the magnetic moment is enhanced below {\it T}$^{\,*}$. This enhancement becomes remarkably large when the pressure is close to {\it P}$_{\rm C}^{\,*}$ = 1.20 GPa. The characteristic temperature  {\it T}$^{\,*}$ depends on both the magnetic field and pressure. Recently Watanabe and Miyake developed a microscopic theory, showing that the growth of magnetization can occur below {\it T}$^{\,*}$ if the characteristic transition at {\it T}$^{\,*}$ corresponds to the coupled CDW/SDW ordering.~\cite{rf:11}  They also reproduced qualitatively the anomalous temperature dependence of $H_{\rm c2}$ by using the field and temperature dependence of {\it T}$^{\,*}$ determined by the present measurement. ~\cite{rf:11}

 On the other hand, there is no experimental proof of the CDW/SDW state. In our neutron scattering experiment, no satellite peaks of the nesting origin was observed along the {\it a}$^{\,*}$- and {\it c}$^{\,*}$- axes. Further diffraction experiments are necessary to survey the satellite peak characteristic to the CDW/SDW state, in which the nesting can occur in any direction.  

   In conclusion, we measured the DC-magnetization of UGe$_2$ under high pressure. The magnetization is remarkably enhanced below {\it T}$^{\,*}$. Especially near the border of {\it P}$_{\rm C}^{\,*}$ = 1.20 GPa, an abrupt increment of magnetization was found around {\it T}$^{\,*}$. We constructed a temperature vs. field phase diagram. 

Å@We are grateful to K. Miyake, K. Machida, S. S. Saxena, H. Yamagami and S Watanabe for helpful discussion. This work was financially supported by the Grant-in-Aid for COE Research (NO.10CE2004) of the Ministry of Education, Science, Sports and Culture, and by CREST, Japan Science and Technology Corporation. One of authors (N. T) was supported by Research Fellowships of the Japan Society for the Promotion of Science for Young Scientists.

\end{document}